\documentclass[twocolumn,aps,prl,tightenlines,floatfix,showpacs]{revtex4}
\usepackage[dvips]{graphicx}
\usepackage{hhline}

\begin{document}

\title{Novel shell-model analysis of the $^{136}$Xe double beta decay nuclear matrix elements }

\author{M. Horoi}
\email{mihai.horoi@cmich.edu}
\affiliation{Department of Physics, Central Michigan University, Mount Pleasant, Michigan 48859, USA}

\author{B.A. Brown}
\email{brown@nscl.msu.edu}
\affiliation{National Superconducting Cyclotron Laboratory,
 and
Department of Physics and Astronomy, Michigan State University, East
Lansing, Michigan 48824, USA}


\begin{abstract}
Neutrinoless 
double beta decay, if observed, could distinguish whether 
neutrino is a Dirac or a Majorana particle, and it could be used to determine the absolute 
scale of the neutrino masses. $^{136}$Xe is one of the most promising
candidates for observing this rare event. However, until recently there were no
positive result for the allowed and less rare two-neutrino double beta decay mode.
The small nuclear matrix element associated with the small half-life represents
a challenge for nuclear structure models used for its calculation.
We report a new shell-model analysis of the two-neutrino double beta decay of
$^{136}$Xe, which takes into account all relevant nuclear orbitals necessary to fully
describe the associated Gamow-Teller strength. We further use the new model to 
analyze the main contributions to the neutrinoless double beta decay half-life, and show that they are also diminished.
\end{abstract}

\maketitle



Neutrinoless double beta ($0 \nu \beta \beta$) decay can only occur by violating the conservation of the total 
lepton number, and if observed it would unravel physics beyond Standard Model (SM) of particle 
physics and it would represent 
a major milestone in the study of the fundamental properties of neutrinos 
\cite{rmp-08}.
Recent results from neutrino oscillation experiments have
demonstrated that neutrinos have mass and they
mix \cite{solar,atmospheric,kamland}. 
In addition, they show that neutrinoless double beta 
decay process could be used to determine the absolute
scale of the neutrino masses, and  can
distinguish whether neutrino is a Dirac or a Majorana particle \cite{sv82}. A
key ingredient for extracting the absolute neutrino masses from
$0\nu\beta\beta$ decay experiments is a precise knowledge of the
nuclear matrix elements (NME) for this process. 
There is a large experimental effort in US and worldwide to investigate the double beta
decay of some even-even nuclei \cite{rmp-08}. Experimental data for
two-neutrino double-beta decay ($2 \nu \beta \beta$) to the ground state (g.s.) 
and excited states already
exist for a group of nuclei \cite{barabash10}. There is no confirmed experimental data so
far for neutrinoless double-beta decay. The prediction, analysis and
interpretation of experimental results, present and expected, are very much
dependent on the precise nuclear structure calculations of corresponding
transition probabilities.

Although many experimental efforts in US and worldwide, such as MAJORANA and GERDA \cite{rmp-08}, 
are pinpointing to the $\beta\beta$ decay of $^{76}$Ge
there are very encouraging results related to the $\beta\beta$ decay of $^{136}$Xe.
For a long time there were  available only upper limits for the $2 \nu \beta \beta$
half-life. Recently, the EXO-200 collaboration reported \cite{exo11,exo12} a precise measurement
of this half life of $2.11\pm0.04(stat)\pm0.21(sys)\times10^{21}$ yr, corresponding to 
a NME of 0.019 MeV$^{-1}$. This large half-life
would imply a smaller background for the associated $0 \nu \beta \beta$ measurement and EXO, 
a larger version of EXO-200 designed for reaching this goal, is under consideration 
\cite{rmp-08}. The upper limit for the $0 \nu \beta \beta$ half-life reported
by EXO-200 is $1.6\times10^{25}$ yr \cite{exo12}.
In addition,
the KamLAND-Zen collaboration reported a $2 \nu \beta \beta$ 
half-life of $2.38\pm0.02(stat)\pm0.14(sys)\times10^{21}$ yr and a upper-limit
for the $0 \nu \beta \beta$ half-life of $5.7\times10^{24}$ yr \cite{zen-prc}.

Since most of the $\beta\beta$ decay emitters are open shell nuclei,
many calculations of the NME have been performed within the pnQRPA approach and its 
extensions 
\cite{sim-97,rodin07,sim-09}.
However, the pnQRPA calculations of the
more common two-neutrino double beta  decay observable, which were measured for about
10 cases \cite{barabash10},
are very sensitive to the variation of the $g_{pp}$ parameter (the strength 
of the particle-particle interactions in the $1^+$ channel) 
\cite{VOG86,SUH88}, 
and this 
drawback persists in spite of various improvements brought by its 
extensions, 
including higher-order QRPA approaches 
\cite{sim-09}. 
 Although the QRPA methods do not seem to be suited to predict the $2\nu\beta\beta$ 
decay half-lives, they
can use the measured $2\nu\beta\beta$ decay half-lives to calibrate the $g_{pp}$ parameters that are further used
to calculate the $0\nu\beta\beta$ decay NME \cite{rodin07}. 
Another method that was recently used to calculate NMEs for most $0\nu\beta\beta$ decay
cases of interest is the Interactive Boson Model (IBA-2) \cite{iba-09}. However, a reliable 
IBA-2 approach for $2 \nu \beta \beta$ decay is not yet available.
  
Recent progress in computer power,  
numerical algorithms, and improved nucleon-nucleon effective interactions, made possible
large-scale configuration-interaction (CI) calculations (also known as shell-model calculations) 
of the $2\nu\beta\beta$ \cite{plb-ca48,caurier-96,retamosa-95,HSB07} 
and $0\nu\beta\beta$ decay
NME \cite{prl100,Hor10}. 
The main advantage of the large-scale shell-model calculations 
is that they take into account all of the many-body correlations for
the orbitals near the Fermi surface.
Also they are also
less dependent on the effective interaction used, as long as these 
are based on realistic nucleon-nucleon interactions with minimal
adjustments to the single-particle energies and some two-body matrix 
elements so they reproduce 
general spectroscopy of the nuclei involved in the decay \cite{Hor10}. 
Their main drawback is the 
limitation imposed by the exploding CI dimensions even for limited increase in the 
size of the valence space used.
The most important success of the large-scale shell-model calculations 
was the correct prediction of the $2\nu\beta\beta$ decay half-life 
for $^{48}$Ca \cite{plb-ca48,exp-ca48}.
In addition, the CI calculations do not have to adjust any additional parameters, i.e.
given the effective interaction and the Gamow-Teller (GT) quenching factor extracted 
from the overall spectroscopy in the respective mass-region, they are able 
to reliably predict the $2\nu\beta\beta$ decay half-life of $^{48}$Ca.


CI methods provide realistic many-body wave
functions (w.f.) for many nuclei from $^{16}$O to $^{100}$Sn and beyond. These wave functions
can describe observables related to specific experiments, e.g. for nuclear astrophysics, and
the electro-weak interactions with the nucleus. 
The minimal valence space
required for $^{136}$Xe involves the $0g_{7/2}1d_{5/2}1d_{3/2}2s_{1/2}0h_{11/2}$ 
orbitals for protons and neutrons  (the
$jj55$ model space). There
are no spurious center-of-mass (CoM) states in the $jj55$ model space
since the CoM operator $\vec{R}$ does not connect any of the orbitals.
The key is to obtain effective interactions (EI) that can provide 
energies and wave functions in $jj55$ model space that
are at a similar level of accuracy as those obtained for the $sd$-shell
\cite{bab06} and for the $pf$-shell \cite{Hon04}. 
The CI $\beta \beta$ decay NME were reported over the years \cite{caurier-96,prl100,caur-136xe}
considering continuous
improvements of the EI.  These calculations
indicate a significant sensitivity of the results to the improving EI. For example,
the quenching factors used to describe $2\nu\beta\beta$ NME varies from 0.74 \cite{caurier-96}
to 0.45 \cite{caur-136xe}, and the $0\nu\beta\beta$ NME varies by a factor of about 3 between
Ref. \cite{caurier-96} and the more recent Ref. \cite{prl100}. One of the drawbacks of model spaces
such as $jj55$ is that in order to maintain center-of-mass purity they do not include all
spin-orbit partners of orbitals such as $0g_{7/2}$ and $0h_{11/2}$. The known effect is that 
the Ikeda sum-rule is not satisfied indicating that some the Gamow-Teller strength, 
which is so important for both types of NME, is missing from this model space. For example,
in $jj55$ typical Ikeda sum-rule for $^{136}$Xe is 52, 
while the expected result is 84 (see also Table \ref{tab1} below). 

In this letter we investigate the effect of extending the model space 
to $jj77$ by including the effects 
of the missing $0g_{9/2}$ and $0h_{9/2}$ orbitals.
The two-body matrix elements with good $  J  $ and $  T  $ were obtained from
the code CENS \cite{cens}. The procedure discussed below
was used to obtain a Hamiltonian for the $jj77$ model space that
we will refer to as $jj77a$.
In the first step,
the short-range part of the N$^{3}$LO potential \cite{n3lo}
was integrated out using the $  V_{{\rm low} k}  $ method \cite{vlowk}. The
relative two-body matrix elements were evaluated in a harmonic-oscillator
basis with $\hbar\omega$=7.874 (a value appropriate for $^{132}$Sn).
In the second step the
interaction was renormalized into the $jj77$ model space assuming a $^{100}$Sn
closed core. The 0g$_{9/2}$ orbital was treated as a hole state, while the
other are treated as particle states. For the energy denominators
we take all orbits in the $jj77$ space to be degenerate with the
other orbitals spaced in units of $\hbar\omega$ above and below.
The core-polarization calculation used the $  \hat{Q}  $-box method and
includes all
non-folded diagrams through second-order in the interaction
and sums up the folded diagrams to infinite order \cite{pol}.
Particle-hole excitations up through 4$\hbar\omega$ were included.
Matrix elements obtained in the proton-neutron basis were transformed
to a good-$  T  $ basis by using the neutron-neutron matrix elements
for the $  T=1  $ components.

The single-particle matrix elements
were obtained starting with the $jj55$ model space for a $^{132}$Sn closed core.
The five single-particle energies for $  0g_{7/2}  $, $  1d_{5/2}  $,
$  1d_{3/2}  $, $  2s_{1/2}  $ and $  0h_{11/2}  $ were adjusted to
reproduce the experimental values for neutron holes related
to the specturm of $^{131}$Sn as given in \cite{sn132g}. The results
obtained for the single-particle energies of protons
related to the spectrum of $^{133}$Sb are in reasonable agreement
with experiment \cite{sn132g} except that the $  1d_{5/2}  $ energy
is too high by
1.2 MeV and the $  1h_{11/2}  $ energy is too high by 2.4 MeV.
Reduction of the diagonal two-body matrix elements by 0.3 MeV for
these two orbitals improves the agreement with experiment
with minimal overall change to the Hamiltonian.
The adjustment of the single-particle energies to experiment
implicitly includes most of the effects due to three-body interactions.

The two-hole spectrum for $^{130}$Sn and the
two-particle spectrum for $^{134}$Te are in best overall agreement
with experiment if the $  T=1  $ matrix elements are multiplied by 0.9.
The results (experiment vs theory) are (1.28, 1.34) MeV for $^{130}$Sn
and (1.22, 1.35) MeV for $^{134}$Te. For application to the larger $jj77$
model space the single-neutron hole energy for $  0g_{9/2}  $ was
placed six MeV below the $ 0g_{7/2}  $ energy in $^{131}$Sn,
and the single-proton particle energy for $  0h_{9/2}  $ was
placed six MeV above the $  0h_{11/2}  $ energy in $^{133}$Sb.

The  $2\nu\beta\beta$ half-life for the  transition from the $0^+$ g.s. of
$^{136}$Xe to the $0^+$ g.s. of $^{136}$Ba can be 
calculated \cite{SC98} using

\begin{equation}
\left[T^{2\nu}_{1/2}\right]^{-1} = 
G^{2\nu} \vert M^{2\nu}_{GT}(0^+)\vert^2 \ ,
\label{hlive}
\end{equation}

\noindent
where $G^{2\nu}$ is a phase space factor that for the
the $2\nu\beta\beta$ of $^{136}$Xe is $1.279\times 10^{-18}\ yr^{-1} MeV^2$ \cite{SC98}, 
$M^{2\nu}_{GT}(0^+)$ is the $2 \nu \beta \beta$ matrix element given by the double Gamow-Teller sum

\begin{equation}
M^{2\nu}_{GT}(0^+) = \sum_k 
\frac{
\langle 0^+_f\vert\vert \sigma\tau^-\vert\vert 1^+_{k}
\rangle 
\langle 1^+_{k}\vert\vert \sigma\tau^-
\vert\vert 0^+_i\rangle
}{E_k + E_0}\ . 
\label{eq1}
\end{equation}

\noindent 
Here $E_k$ is the excitation energy of the $1^+_k$ state of $^{136}$Cs
and $E_0 = \frac{1}{2}Q_{\beta\beta}(0^+) + \Delta M = 1.31$ MeV,  
where we used the recently reported \cite{qbbxe} Q-value $Q_{\beta\beta}(0^+)=2.458$ MeV  
corresponding to the $\beta\beta$ 
decays to the g.s. of $^{136}$Ba;  $\Delta M$ is the 
$^{136}$Cs - $^{136}$Xe mass difference.  

\begin{table}[tbe]
	\caption{Matrix elements in MeV$^{-1}$ for $2\nu$ decay 
calculated using the standard  quenching factor 0.74 for the Gamow-Teller operator using
different number of excitations from $jj55$ to the larger model space. 
Last column indicate the calculated Ikeda sum-rule for $^{136}$Xe. } 
	\begin{center}
        \begin{tabular}{|c|c|c|c|c|c|}
            \hline
$n\ (0^+)$ & $n\ (1^+)$ & $M^{2\nu}$ & Ikeda \\
	\hline
 0 & 0 & 0.062 & 52 \\
 0 & 1 & 0.091 & 84 \\
 1 & 1 & 0.037 & 84 \\
 1 & 2 & 0.020 & 84\\
\hline
      
        \end{tabular}
	\end{center}
\label{tab1}
\end{table}

In Ref. \cite{HSB07} we fully diagonalized 250 $1^+$ states in the intermediate nucleus to calculate
the $2 \nu \beta \beta$ decay NME for $^{48}$Ca. This procedure can be used for somewhat heavier nuclei
using the J-scheme shell-model code NuShellX \cite{nushellx}, 
but for cases with large dimension one needs an alternative
method. 
Here we used a novel improvement \cite{mh-cssp10} of the known strength-function approach
\cite{plb-ca48}, which 
is very efficient for large model cases. 
such as $jj55$ and $jj77$. 
For example, to calculate the NME for the decays of $^{128}$Te   in $jj55$  and $^{136}$Xe 
 in $jj77$ ($n=1$ for $0^+$ and $n=2$ for $1^+$ in Table \ref{tab1}) one needs
to solve problems with m-scheme dimensions of up the order of to ten billion.

The result when restricting the $jj77$ model space to $jj55$ is given on the first line
in Table \ref{tab1}. As already mentioned, the Ikeda sum-rule is only 52 rather then
84, indicating that not all GT strength is available in the $jj55$ space.
Although the excitation energies of the GT strength
distribution are reasonably well reproduced, the GT operator $\sigma \tau$ has to be multiplied 
by a quenching factor due to correlations beyond the $jj77$ model space.
In typical one major-shell calculations, such as the $sd$ or $pf$, this quenching factor
was determined to be around 0.74-0.77 (see e.g. Ref. \cite{sdgt,pfgt}) that
is consistent with that obtained in second-order perturbation theory \cite{arima,towner}.
Here we use 0.74. 
Ref. \cite{caur-136xe} suggests
that one should use a lower quenching factor in the $jj55$ model space, 0.45, 
to get an NME consistent with
the recent experimental data. Indeed, our matrix elements 
in the $jj55$ model space becomes 0.022 MeV$^{-1}$ when
0.45 is used. 

However, it would be important to check if the missing spin-orbit partners are
responsible for the larger result; the relative phases in Eq. (\ref{eq1}) could lead
to large cancellations. Here we consider the larger $jj77$ model space, but we could
only allowed few $n$ particle being excited from the $0g_{9/2}$ orbital or to
the $h_{9/2}$ orbital, relative to $jj55$. 
Table \ref{tab1} also presents the NME for different combinations
of the allowed $n$ for the initial and final $0^+$ states and the intermediate $1^+$
states. One can see that when $n$ is 1 for the $0^+$ states and 2 for the $1^+$
states the NME decreases almost to the experimental value without the need of artificially
reducing the quenching factor. In addition, the Ikeda sum-rule is always satisfied
in the larger model space.

One should mention that in the $jj77$ model space the wave functions could have CoM spurious
components. We checked our initial and final $0^+$ g.s. w.f. and we found negligible (less
than 3 keV) spurious contribution to expectation values of the CoM Hamiltonian.
We did not check the amount of CoM spuriously in the intermediate $1^+$ states, but it's
unlikely to be large because the strength function method \cite{mh-cssp10} performs 
a small number of Lanczos iterations (about 30) starting with a door-way state obtained by applying 
the GT operator on the largely nonspuroius $0^+$ state.
As a further check we compared the GT strength (BGT) for the transition from the g.s.
of $^{136}$Xe to the first $1^+$ state in $^{136}$Cs with recent experimental data
\cite{GT11}. 
Table I of Ref. \cite{GT11} provides a BGT of 0.149(21) for the first $1^+$ state at 0.59 MeV, 
but we learned
\cite{Dieter} that this will be updated to 0.24(7). Our BGT is 0.51 in the $jj55$ model space,
but 0.34 in the largest $jj77$ model space, much closer to the experimental value.
Although, we cannot verify if the calculations are converged
we can conclude that including all spin-orbit partners 
is essential 
for a good description of the $2 \nu \beta \beta$ for $^{136}$Xe.

\begin{table}[tbe]
	\caption{Matrix elements  for $0\nu$ decay using  two SRC models \cite{sim-09}, CD-Bonn (SRC1) 
	and Argonne (SRC2). The  upper values of the neutrino physics parameters $\eta^{up}_j$ in units 
	of $10^{-7}$ are calculated using the $G^{0\nu}$ from Refs. \cite{SC98} and \cite{kipf12}.} 
	\begin{center}
        \begin{tabular}{|c|c|c|c|c|c|}
            \hline
 \multicolumn{2}{|c|}{} & & & & \\
  \multicolumn{2}{|c|}{} & $M^{0 \nu}_{\nu}$ & $M^{0 \nu}_N$ & $M^{0 \nu}_{\lambda'}$
& $M^{0 \nu}_{\tilde{q}}$  \\
\hline
$n=0$ & SRC1 &  2.21 & 143.0 & 1106. & 206.8 \\
 & SRC2 & 2.06 & 98.79 & 849.0 & 197.2 \\    
\hline      
$n=1$ & SRC1 &  1.46 & 128.0 & 1007 & 157.8 \\
  \cline{2-6}
 & $\left| \eta^{up}_j \right|$ \cite{SC98} & 8.19 & 0.093 & 0.012 & 0.075 \\
& $\left| \eta^{up}_j \right|$ \cite{kipf12} & 9.02 & 0.103 & 0.013 & 0.083 \\
            \hline
        \end{tabular}
	\end{center}
\label{tab2}
\end{table}

Having tuned our nuclear
structure techniques to getting an accurate description of the two-neutrino double-beta decay
we  calculate the  NME necessary for the analysis
of the neutrinoless double-beta decay half-life $^{136}$Xe \cite{Hor10,ca48-12}. 
Considering the most important mechanisms that could be responsible for 
$0\nu\beta\beta$ decay \cite{ves12} one can write the $0\nu\beta\beta$ half-life

\begin{widetext}
\begin{eqnarray}
\left[ T^{0\nu}_{1/2} \right]^{-1}   =  G^{0\nu}\left| \eta_{\nu L} M^{0 \nu}_{\nu} 
 +  \eta_{N} M^{0 \nu}_N 
 +  \eta_{\lambda'} M^{0 \nu}_{\lambda'} 
 +  \eta_{\tilde{q}} M^{0 \nu}_{\tilde{q}} \right|^2,
\label{t0red}
\end{eqnarray}
\end{widetext}

\noindent
where $M^{0\nu}_j$ NME and $\eta_j$  neutrino
physics parameters for light neutrino exchange ($j=\nu$), heavy neutrino exchange ($j=N$),
gluino exchange ($j=\lambda'$) and squark-neutrino mechanism ($j=N$)
are described in Refs. \cite{ca48-12,ves12}. $G^{0\nu}$ is a phase space factor
 tabulated in several publications. One widely used value
 \cite{SC98} is $43.7\times10^{-15}$ yr$^{-1}$. 
A recent publication \cite{kipf12} proposes 
$36.05\times10^{-15}$ yr$^{-1}$, which is about  20\% lower.
The results for the NME calculated in the closure approximation are presented in Table \ref{tab2} using the $n=0$ and $n=1$ $0^+$ w.f. (see Table I).
Two recent short-range correlations (SRC) parametrizations
are used \cite{sim-09,Hor10}. No quenching of the bare transition operator was used 
\cite{Hor10,eh-09}.
The $M^{0\nu}_\nu$ for the $jj55$ model space ($n=0$) is consistent with 
other recent shell-model results \cite{prl100}. 
The NME for the other three mechanisms calculated
within a shell-model approach are reported
here for the first time. The NME in the largest space ($n=1$) are 10-30\% lower.
These results suggest that the inclusion of the spin-orbit partners, which proved to be 
significant for a good description of the $2\nu\beta\beta$ NME, could be also very important
for a reliable description of the $0\nu\beta\beta$ NME. In addition, they indicate that
the net effect is a decrease of the NME rather than an increase (an assumption often used
to understand the lower shell-model value relative to the results of other methods, such 
as QRPA, IBA-2, Projected Hatree-Fock Bogoliubov 
\cite{phfb}, and
 Generator Coordinate Method 
\cite{gcm}).
Table \ref{tab2} also presents upper limits for the neutrino physics parameters 
$\left| \eta^{up}_j \right|$
under the assumption of single mechanism dominance.
They were obtained from Eq. (\ref{t0red}) using the lower limit for the half-life 
$1.6\times10^{25}$ yr from Ref. \cite{exo12} and the two phase space factors of
Refs. \cite{SC98} and \cite{kipf12}. Using the upper limits for 
$\left| \eta_{\nu L} \right| = m_{\beta\beta}/m_e$ 
one can extract an upper limit for the effective neutrino
mass $m_{\beta\beta}$ of 0.42-0.46 eV.

In conclusion, we reported a new shell-model analysis of the two-neutrino double 
beta decay of
$^{136}$Xe that takes into account all relevant nuclear orbitals necessary for a good
description of the Gamow-Teller strength. We show that this extension of the valence  
space can account for the small NME without recourse to an artificially small quenching factor.
We also show that it could lead to smaller NME for the most interesting neutrinoless 
double beta decay mode.

\vspace{0.3cm}
Support from U.S. NSF Grant PHY-1068217 is acknowledged. 
M.H. acknowledges the SciDAC Grant NUCLEI.


\end{document}